\documentclass[twopage,11pt] {article}

\usepackage{psfig}

\begin{document}

\title{Accretion-driven millisecond X-ray pulsars}
\author{Rudy Wijnands}         

\maketitle

\pagestyle{myheadings} 
\thispagestyle{plain}         
\markboth{Rudy Wijnands}{Accretion-driven millisecond X-ray pulsars} 
\setcounter{page}{1}

\noindent
{\bf Abstract:} I present an overview of our current observational
knowledge of the six known accretion-driven millisecond X-ray
pulsars. A prominent place in this review is given to SAX
J1808.4--3658; it was the first such system discovered and currently
four outbursts have been observed from this source, three of which
have been studied in detail using the {\it Rossi X-ray Timing
Explorer} satellite. This makes SAX J1808.4--3658 the best studied
example of an accretion-driven millisecond pulsar. Its most recent
outburst in October 2002 is of particular interest because of the
discovery of two simultaneous kilohertz quasi-periodic oscillations
and nearly coherent oscillations during type-I X-ray bursts. This is
the first (and so far only) time that such phenomena are observed in a
system for which the neutron star spin frequency is exactly known. The
other five systems were discovered within the last three years (with
IGR J00291+5934 only discovered in December 2004) and only limited
results have been published.

\section{Introduction \label{section:intro}}

Ordinary pulsars are born as highly-magnetized (B $\sim 10^{12}$ G),
rapidly rotating (P $\sim$ 10 ms) neutron stars which spin down on
timescales of 10 to 100 million years due to magnetic dipole
radiation.  However, a number of millisecond (P $<$ 10 ms) radio
pulsars is known with ages of billions of years and weak (B $\sim
10^{8-9}$ G) surface magnetic fields.  Since many of these millisecond
pulsars are in binaries, it has long been suspected (see, e.g.,
Bhattacharya \& van den Heuvel 1991 for an extended review) that the
neutron stars were spun up by mass transfer from a stellar companion
in a low-mass X-ray binary (LMXB), but years of searching for coherent
millisecond pulsations in LMXBs failed to yield a detection (Vaughan
et al.~1994 and references therein).  The launch of the NASA {\it
Rossi X-ray Timing Explorer} ({\it RXTE}) brought the discovery of
kilohertz quasi-periodic oscillations (kHz QPOs; Strohmayer et
al.~1996; Van der Klis et al.~1996) as well as nearly coherent
oscillations ('burst oscillations') during type-I X-ray bursts in a
number of LMXBs (e.g., Strohmayer et al.~1996), providing
tantalizingly suggestive evidence for weakly magnetic neutron stars
with millisecond spin periods (see Van der Klis 2000, 2004 and
Strohmayer \& Bildsten 2003 for more details about kHz QPOs and burst
oscillations in LMXBs).

In April 1998 the first accretion-driven millisecond X-ray pulsar (SAX
J1808.4--3658) was discovered (Wijnands \& van der Klis 1998a) proving
that indeed neutron stars in LMXBs can spin very rapidly. This
conclusion was further strengthened by the discovery of four
additional systems in 2002 and 2003 (Markwardt et al.~2002a, 2003a,
2003b, Galloway et al.~2002), and recently, in December 2004, with the
discovery of IGR J00291+5934 as a millisecond X-ray pulsar (Markwardt
et al.~2004a). Here, I will give a brief summary of our current
observational knowledge of those accretion-driven millisecond X-ray
pulsars. Preliminary versions of this review were published by
Wijnands (2004a, 2004b).

\section{SAX J1808.4--3658
\label{section:1808}}

\begin{figure}[t]
\begin{center}
\begin{tabular}{c}
\psfig{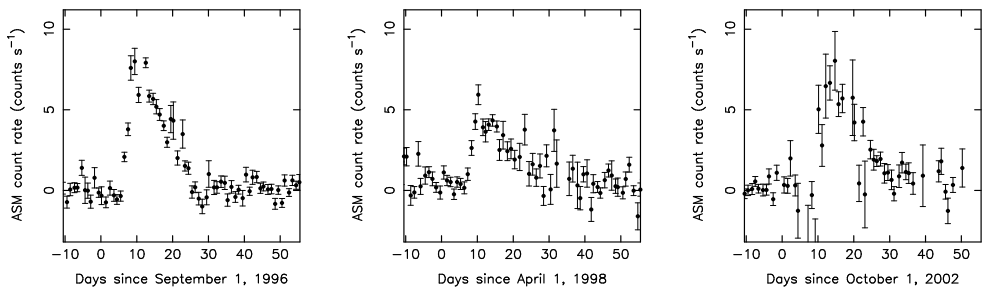}
\end{tabular}
\caption{
The {\it RXTE}/ASM light curves of SAX J1808.4--3658 during the
September 1996 outburst (left), the April 1998 outburst (middle) and
the October 2002 outburst (right).  These light curves were made using
the public ASM data available at http://xte.mit.edu/ASM\_lc.html. The
count rates are for the 2--12 keV energy range and are daily
averages.
\label{fig:1808_asm} }
\end{center}
\end{figure}

\subsection{The September 1996 Outburst
\label{subsection:1996outburst}}

In September 1996, a new X-ray transient and LMXB was discovered with
the Wide Field Cameras (WFCs) aboard the Dutch-Italian {\it BeppoSAX}
satellite and the source was designated SAX J1808.4--3658 (In 't Zand
et al.~1998). Three type-I X-ray bursts were detected, demonstrating
that the compact object in this system is a neutron star. From those
bursts, a distance estimate of 2.5 kpc was determined (In 't Zand et
al.~1998, 2001). The maximum luminosity during this outburst was $\sim
10^{36}$ erg s$^{-1}$, significantly lower than the peak outburst
luminosity of 'classical' neutron star transients (which typically can
reach a luminosity of $10^{37}$ to $10^{38}$ ergs s$^{-1}$). This low
peak luminosity showed that the source was part of the growing group
of faint neutron-star X-ray transients (Heise et al.~1999).  The
outburst continued for about three weeks (see Fig.~\ref{fig:1808_asm})
after which the source was thought to have returned to quiescence.
However, it was found (Revnivtsev 2003) that the source was detected
on October 29, 1996 (using slew data obtained with the proportional
counter array [PCA] aboard {\it RXTE}) with a luminosity of about a
tenth of the outburst peak luminosity. This demonstrates that six
weeks after the main outburst the source was still active (possibly
only sporadically), which might indicate that at the end of this
outburst the source behaved in a manner very similar to what was seen
during its 2000 and 2002 outbursts (see
\S~\ref{subsection:2000outburst} and
\S~\ref{subsection:2002outburst}).

After it was found that SAX J1808.4--3658 harbors a millisecond pulsar
(\S~\ref{subsection:1998outburst}), the three observed X-ray bursts
seen with {\it BeppoSAX}/WFC were scrutinized for potential burst
oscillations (In 't Zand et al.~2001). A marginal detection of a 401
Hz oscillation was made in the third burst. This result suggested that
the burst oscillations observed in the other, non-pulsating,
neutron-star LMXBs occur indeed at their neutron-star spin
frequencies. This result has been confirmed by the recent detection of
burst oscillations during the 2002 outburst of SAX J1808.4--3658
(\S~\ref{subsubsection:bursts}).

\begin{figure}[t]
\begin{center}
\begin{tabular}{c}
\psfig{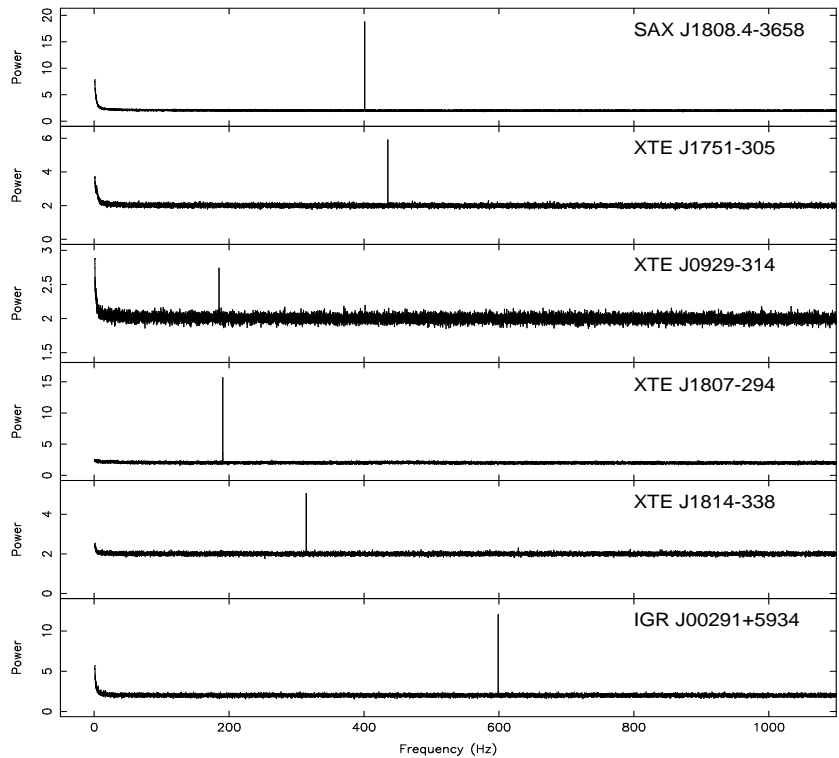}
\end{tabular}
\caption{
Examples of power spectra for each of the six currently known
millisecond X-ray pulsars showing the pulsar spikes.
\label{fig:pulsations} }
\end{center}
\end{figure}

\subsection{The April 1998 outburst
\label{subsection:1998outburst}}
 
On April 9, 1998, {\it RXTE}/PCA slew observations indicated that SAX
J1808.4--3658 was active again (Marshall 1998; see
Fig.~\ref{fig:1808_asm} for the {\it RXTE}/ASM light curve during this
outburst). Using public TOO observations of this source from April 11,
it was discovered (Wijnands \& Van der Klis 1998a) that coherent 401
Hz pulsations (Fig.~\ref{fig:pulsations}) were present in the
persistent X-ray flux of the source, making it the first
accretion-driven millisecond X-ray pulsar discovered. After this
discovery, several more public {\it RXTE} observations were made
(using the PCA) which were used by several groups to study different
aspects of the source. I will briefly mention those results and I
point to references for the details.

A detailed analysis of the coherent timing behavior (Chakrabarty \&
Morgan 1998) showed that the neutron star was in a tight binary with a
very low-mass companion star in a $\sim$2-hr orbital period. Due to
the limited amount of data obtained during this outburst, only an
upper limit of $<7 \times 10^{-13}$ Hz s$^{-1}$ could be obtained on
the pulse-frequency derivative (Chakrabarty \& Morgan 1998). Studies
of the X-ray spectrum (Gilfanov et al.~1998; Heindl \& Smith 1998; see
also Gierlinski et al.~2002 and Poutanen \& Gierlinski 2003) and the
aperiodic rapid X-ray variability (Wijnands \& van der Klis 1998b; see
also Van Straaten et al.~2005) showed an object that, apart from its
pulsations, is remarkably similar to other LMXBs with comparable
luminosities (the atoll sources). There is apparent modulation of the
X-ray intensity at the orbital period, with a broad minimum when the
pulsar is behind the companion (Chakrabarty \& Morgan 1998; Heindl \&
Smith 1998). Cui et al.~(1998) and Ford (2000) reported on the
harmonic content, energy dependency, and soft phase lag of the
pulsations. The main result of those studies is that the low-energy
pulsations lag the high-energy ones by as much as $\sim$200 $\mu$s
($\sim$8\% of the pulsation period; see Cui et al.~[1998], Ford
[2000], and Poutanen \& Gierlinski [2003] for possible explanations
for these soft lags).

Another interesting aspect is that the source first showed a steady
decline in X-ray flux, which after ~2 weeks suddenly accelerated
(Gilfanov et al.~1998; Cui et al.~1998;
Fig.~\ref{fig:1808_1998v2002}). This behavior has been attributed to
the fact that the source might have entered the 'propeller regime' in
which the accretion is centrifugally inhibited (Gilfanov et
al.~1998). However, after the onset of the steep decline the
pulsations could still be detected (Cui et al.~1998) making this
interpretation doubtful. A week after the onset of this steep decline,
the X-ray flux leveled off (Cui et al.~1998; Wang et al.~2001), but as
no further {\it RXTE}/PCA observations were made, the X-ray behavior
of the source at the end of the outburst remained unclear. The source
might have displayed a similar long-term episode of low-luminosity
activity as seen at the end of its 2000 and 2002 outbursts (see
\S~\ref{subsection:2000outburst} and
\S~\ref{subsection:2002outburst}).

SAX J1808.4--3658 was not only detected and studied in X-rays but also
in the optical, IR, and in radio bands. The optical/IR counterpart of
SAX J1808.4--3658 (later named V4580 Sgr; Kazarovets et al.~2000) was
first discovered by Roche et al.~(1998) and subsequently confirmed by
Giles et al.~(1998). A detailed study of the optical behavior during
this outburst was reported by Giles et al.~(1999) and Wang et
al.~(2001). Both papers reported that the peak V magnitude of the
source was $\sim$16.7 and the source decayed in brightness as the
outburst progressed. The brightness of the source leveled off at
around V $\sim$ 18.5 (I $\sim$ 17.9) about $\sim$2 weeks after the
peak of the outburst. It stayed at this level for at least several
weeks before it further decreased in brightness. This behavior
suggests that the source was indeed still active for a long period
after the main outburst.

It was also reported (Giles et al.~1999) that the optical flux was
modulated at the 2-hr orbital period of the system. Modeling the X-ray
and optical emission from the system using an X-ray-heated accretion
disk model yielded a Av of 0.68 and an inclination of cos i = 0.65
(Wang et al.~2001), resulting in a mass of the companion star of
0.05--0.10 solar masses. During some of the IR observations, the
source was too bright to be consistent with emission from the disk or
the companion star, even when considering X-ray heating. This IR
excess might be due to synchrotron processes, likely related to an
outflow or ejection of matter (Wang et al.~2001). Such an ejection
event was also confirmed by the discovery of the radio counterpart
(Gaensler et al.~1999). The source was detected with a 4.8 GHz flux of
$\sim$0.8 mJy on 1998 April 27, but it was not detected at earlier or
later epochs.

\subsection{The January 2000 outburst
\label{subsection:2000outburst}}

\begin{figure}[t]
\begin{center}
\begin{tabular}{c}
\psfig{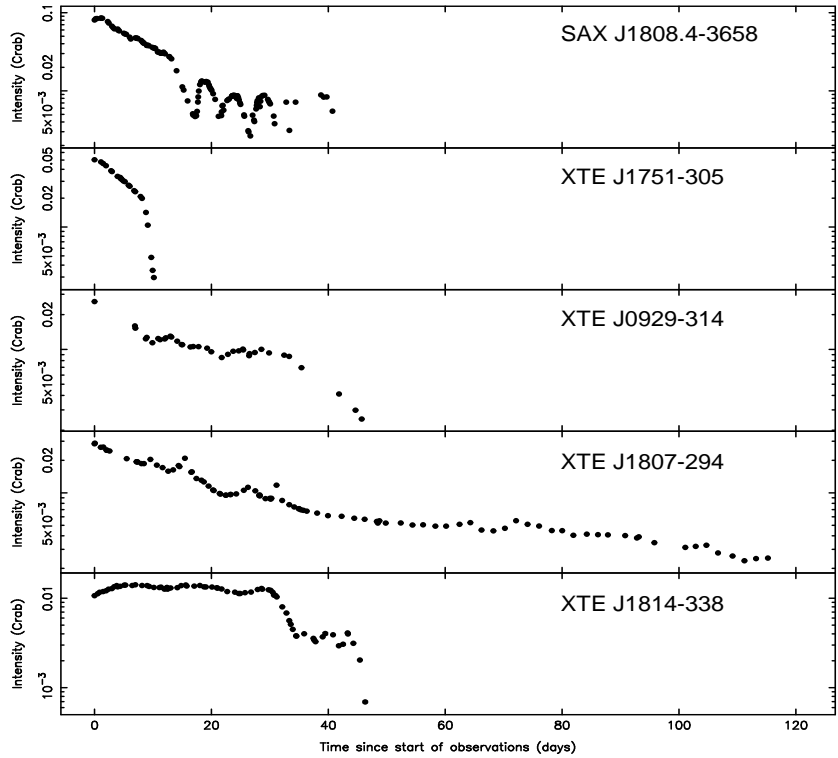}
\end{tabular}
\caption{
The {\it RXTE}/PCA light curves of five of the six accretion-driven
millisecond X-ray pulsars. The data for SAX J1808.4-3658 was obtained
during its 2002 outburst. The data were taken from van Straaten et
al.~(2005), except for XTE J1807--294 which were taken from Linares et
al.~(2005 in preparation).
\label{fig:lc} }
\end{center}
\end{figure}

On January 21, 2000, SAX J1808.4--3658 was again detected (Wijnands et
al.~2001) with the {\it RXTE}/PCA at a flux level of $\sim$10--15
mCrab (2--10 keV), i.e. about a tenth of the peak fluxes observed
during the two previous outbursts. Using follow-up {\it RXTE}/PCA
observations, it was found that the source exhibited low-level
activity for several months (Wijnands et al.~2001). Due to solar
constraints the source could not be observed before January 21 but
likely a true outburst occurred before that date and we only observed
the end stages of this outburst. This is supported by the very similar
behavior of the source observed near the end of its 2002 October
outburst (see \S~\ref{subsection:2002outburst};
Fig.~\ref{fig:lc}).

\begin{figure}[t]
\begin{center}
\begin{tabular}{c}
\psfig{figure=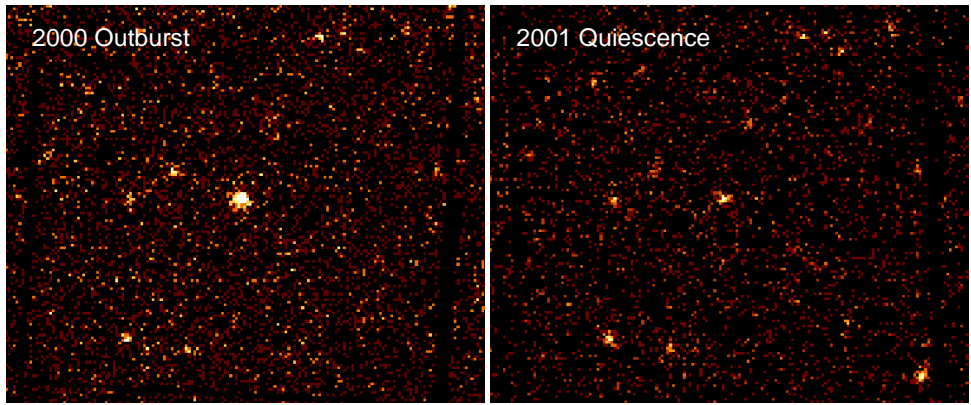,width=13cm}
\end{tabular}
\caption{ 
The {\it XMM-Newton} images of the field containing SAX J1808.4--3658
during its 2000 outburst (left panel; Wijnands 2003) and when the
source was in quiescence (in 2001; right panel; see Campana et
al.~2002). Clearly, SAX J1808.4--3658 (the source in the middle of the
image) was brighter (albeit if only by a factor of a few) during the
2000 outburst observation than during the quiescent observation.
\label{fig:XMM} }
\end{center}
\end{figure}

During the 2000 outburst, SAX J1808.4--3658 was observed (using {\it
RXTE}/PCA) on some occasions at luminosities of ~$\sim$$10^{35}$ ergs
s$^{-1}$, but on other occasions (a few days earlier or later) it had
luminosities of $\sim$$10^{32}$ ergs s$^{-1}$ (as seen during {\it
BeppoSAX} and {\it XMM-Newton} observations; Wijnands et al.~2002,
Wijnands 2003; see Fig.~\ref{fig:XMM} left panel). This demonstrates
that the source exhibited extreme luminosity swings (a factor of
$>$1000) on timescales of days. During the {\it RXTE} observations, it
was also found that on several occasions the source exhibited strong
(up to 100 \% r.m.s. amplitude) violent flaring behavior with a
repetition frequency of about 1 Hz (Van der Klis et al.~2000;
Fig.~\ref{fig:1hz}). During this episode of low-level activity, the
pulsations at 401 Hz were also detected.

\begin{figure}[t]
\begin{center}
\begin{tabular}{c}
\psfig{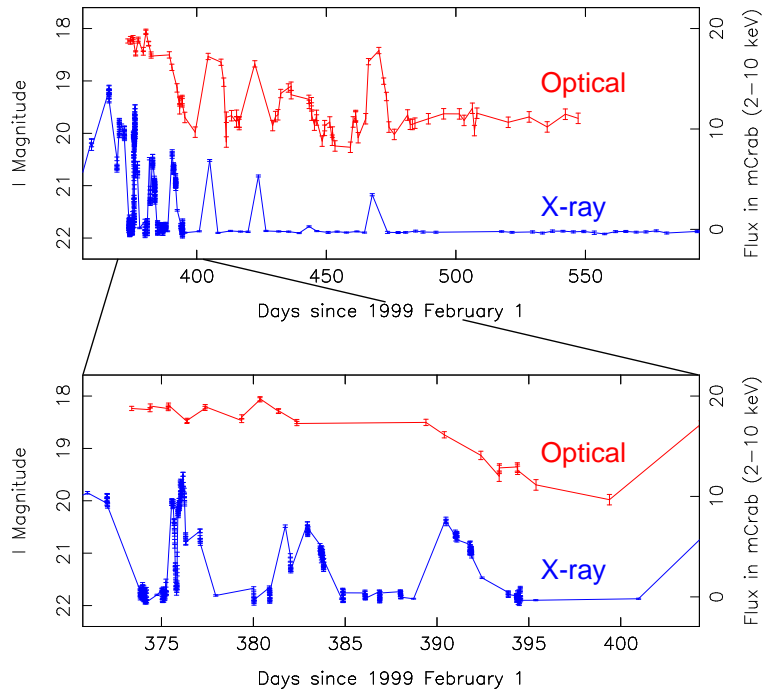}
\end{tabular}
\caption{
The {\it RXTE}/PCA (Wijnands et al.~2001) and the optical (I band)
light curves (Wachter et al.~2000) of SAX J1808.4--3658 as observed
during its 2000 outburst. The optical data were kindly provided by
Stefanie Wachter.
\label{fig:2000_lc} }
\end{center}
\end{figure}

The source was again detected in optical, albeit at a lower brightness
than during the 1998 outburst (Wachter \& Hoard 2000). This is
consistent with the lower X-ray activity seen for the source. The
source was frequently observed during this outburst and preliminary
results were presented by Wachter et al.~(2000). The main results are
presented in Figure~\ref{fig:2000_lc} (reproduced with permission from
Stefanie Wachter). The optical and X-ray brightness of the source are
correlated at the end of the outburst, although one optical flare
(around day 435--440 in Fig.~\ref{fig:2000_lc}) was not accompanied by
an X-ray flare. However, the optical and X-ray observations were not
simultaneous, which means that a brief (around a few days) X-ray flare
could have been missed. During the earlier stages of the outburst, the
X-ray and the optical behavior of the source were not correlated
(Fig.~\ref{fig:2000_lc} lower panel): the source is highly variable in
X-rays, but quite stable in optical with only low amplitude
variations. This stable period in the optical is very similar to the
episode of stable optical emission in the late stages of the 1998
outburst, suggesting this is typical behavior for this source.

\subsection{The October 2002 outburst
\label{subsection:2002outburst}}

In 2002 October, the fourth outburst of SAX J1808.4--3658 was detected
(Markwardt et al.~2002b), immediately launching an extensive {\it
RXTE}/PCA observing campaign. The main results are summarized below.

\subsubsection{The X-ray light curve
\label{subsubsection:lightcurve}}

The {\it RXTE}/PCA light curve for this outburst is shown in
Figure~\ref{fig:lc} (see Fig.~\ref{fig:1808_asm} for the
ASM light curve). During the first few weeks, the source decayed
steadily, until the rate of decline suddenly increased, in a manner
similar to what was observed during the 1998 outburst (see
\S~\ref{subsection:1998outburst}). During both the 1998 and 2002
outbursts, the moment of acceleration of the decline occurred at about
two weeks after the peak of the outburst. Approximately five days
later the X-ray count rate rapidly increased again until it reached a
peak of about a tenth of the outburst maximum. After that the source
entered a state in which the count rate rapidly fluctuated on time
scales of days to hours, very similar to the 2000 low-level activity
(see \S~\ref{subsection:2000outburst}). The 2002 outburst light curve
is the most detailed one seen for this source and it exhibits all
features seen during the previous three outbursts of the source (the
initial decline, the increase in the decline rate, the long-term
low-level activity), demonstrating that this behavior is typical for
this source.

\subsubsection{The X-ray bursts and the burst oscillations
\label{subsubsection:bursts}}

During the first five days of the outburst, four type-I X-ray bursts
were detected. Burst oscillations were observed during the rise and
decay of each burst, but not during the peak (Chakrabarty et
al.~2003). The frequency in the burst tails was constant and identical
to the spin frequency, while the oscillation in the burst rise showed
evidence for a very rapid frequency drift of up to 5 Hz. This
frequency behavior and the absence of oscillations at the peak of the
bursts is similar to the burst oscillations seen in other,
non-pulsating neutron star LMXBs, demonstrating that indeed the
burst-oscillations occur at the neutron-star spin frequency in all
sources. As a consequence, the spin frequency is now known for 18
LMXBs (12 burst-oscillations sources and 6 pulsars) with the highest
spin frequency being 619 Hz. The sample of burst-oscillation sources
was used to demonstrate that neutron stars in LMXBs spin well below
the break-up frequency for neutron stars. This could suggest that the
neutron stars are limited in their spin frequencies, possible due to
the emission of gravitational radiation (Chakrabarty et al.~2003;
Chakrabarty 2004).

\begin{figure}[t]
\begin{center}
\begin{tabular}{c}
\psfig{figure=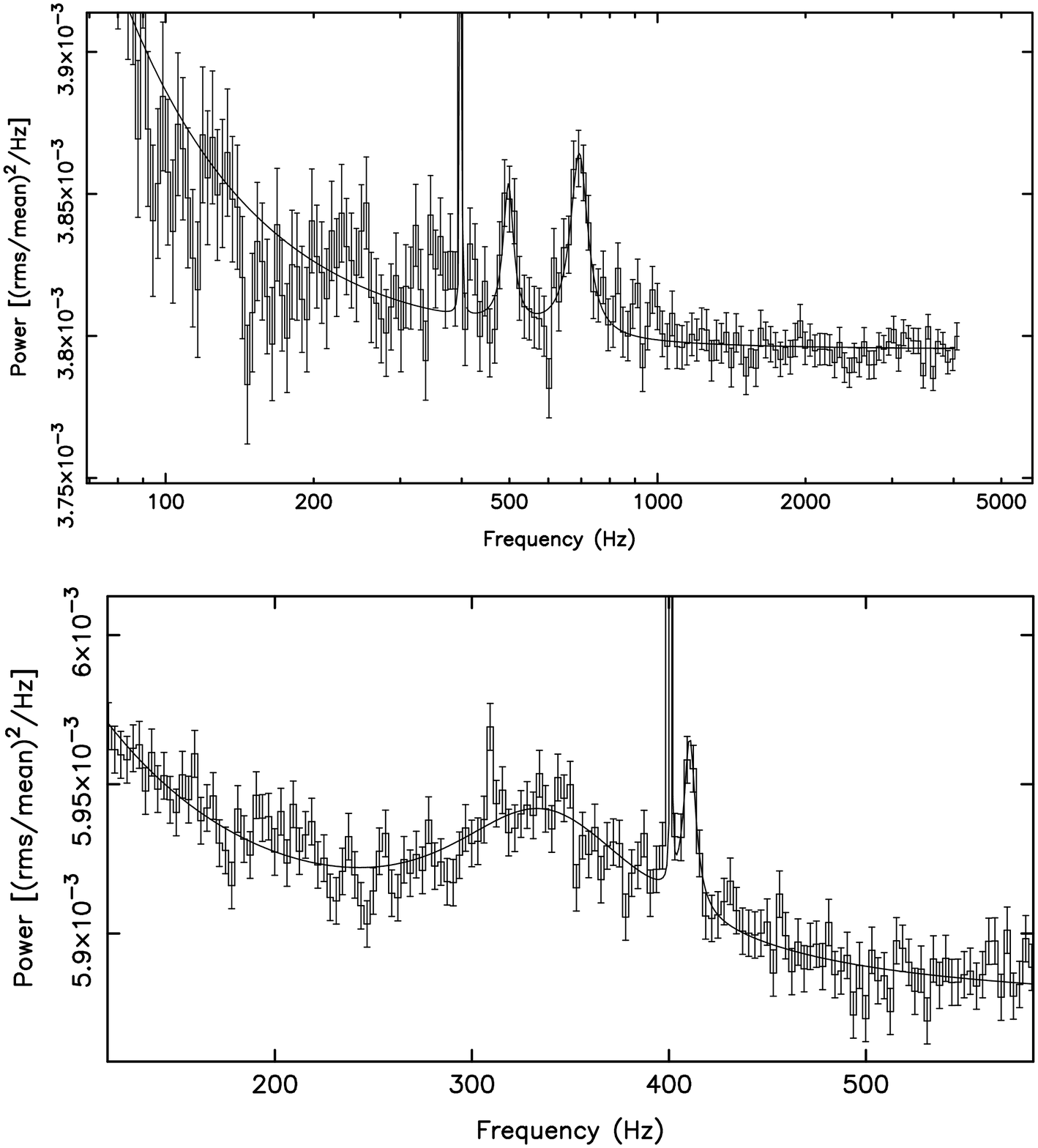,height=10cm,width=11cm}
\end{tabular}
\caption{ 
The power spectrum of SAX J1808.4--3658. The top panel shows the
two simultaneous  kHz QPOs discovered during its 2002 outburst.
The bottom panel shows the enigmatic 410 Hz QPO also seen during
this outburst. The figures are adapted from Wijnands et al.~(2003).
\label{fig:khzqpos_1808} }
\end{center}
\end{figure}

\subsubsection{The kHz QPOs
\label{subsubsection:khzqpos}}

Wijnands et al.~(2003) reported on the discovery of two simultaneous
kHz QPOs during the peak of the outburst with frequencies of $\sim$700
and $\sim$500 Hz (Fig.~\ref{fig:khzqpos_1808} top panel). This was the
first detection of twin kHz QPOs in a source with a known
spin-frequency. The frequency separation of those two kHz QPOs is only
$\sim$200 Hz, significantly below the 401 Hz expected in the
beat-frequency models proposed to explain the kHz QPOs. Therefore,
those models are falsified by the discovery of kHz QPOs in SAX
J1808.4--3658. The fact that the peak separation is approximately half
the spin frequency suggests that the kHz QPOs are indeed connected to
the neutron-star spin frequency, albeit in a way not predicted by any
existing model at the time of the discovery. The lower-frequency kHz
QPO was only seen during the peak of the outburst (October 16, 2002)
but the higher-frequency kHz QPO could be traced throughout the main
part of the outburst (Wijnands et al.~2003). In addition to the twin
kHz QPOs, a third kHz QPO was found with frequencies ($\sim$410 Hz)
just exceeding the pulse frequency (Fig.~\ref{fig:khzqpos_1808} bottom
panel; Wijnands et al.~2003). The nature of this QPO is unclear but it
might be related to the side-band kHz QPO seen in several other
sources (Jonker et al.~2000).

Wijnands et al.~(2003) pointed out that there appear to exist two
classes of neutron-star LMXBs: the 'fast' and the 'slow' rotators. The
fast rotators have spin frequencies $>$$\sim$400 Hz and the frequency
separation between the kHz QPOs is roughly equal to half the spin
frequency. In contrast, the slow rotators have spin frequencies below
$<$$\sim$400 Hz and a frequency separation roughly equal to the spin
frequency. These latest kHz QPO results have spurred new theoretical
investigations into the nature of kHz QPO, involving spin induced
resonance in the disk (e.g., Wijnands et al.~2003; Kluzniak et
al.~2004; Lee et al.~2004; Lamb \& Miller 2004; Kato 2004).

\begin{figure}[t]
\begin{center}
\begin{tabular}{c}
\psfig{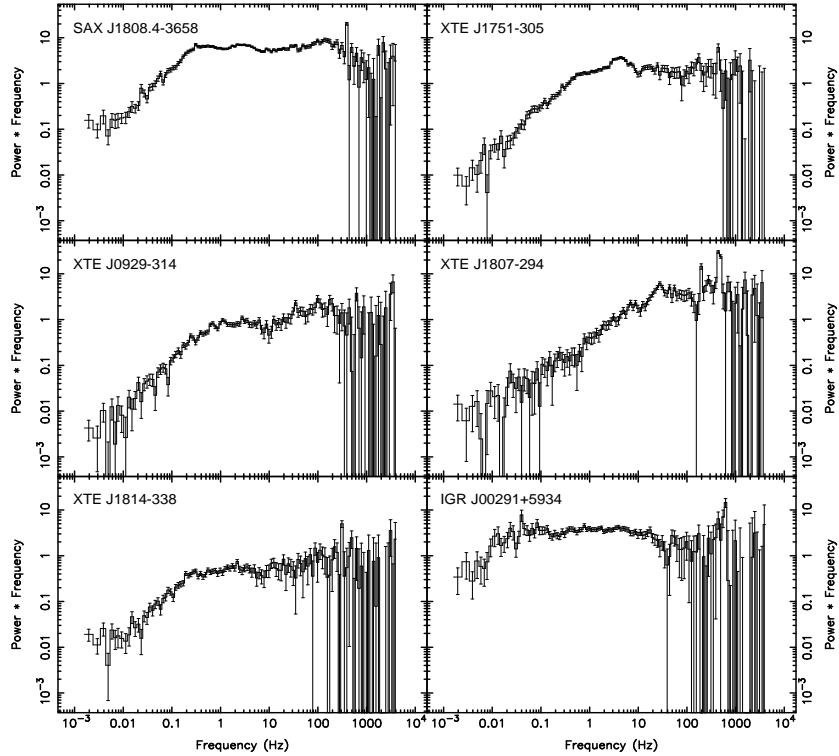}
\end{tabular}
\caption{ 
Examples of the aperiodic timing features seen in the six millisecond
pulsars. For SAX J1808.4--3658 we show a power spectrum obtained
during its 1998 outburst.
\label{fig:lfqpos} }
\end{center}
\end{figure}

\subsubsection{The low-frequency QPOs
\label{subsubsection:lfqpos}}

During the peak of the outburst and in its subsequent decay,
broad-noise and QPOs with frequencies between 10 and 80 Hz were
detected in the power spectra (Fig.~\ref{fig:lfqpos}). Similar
phenomena have been observed in other non-pulsating systems and are
likely to be related to the noise components seen in SAX
J1808.4--3658. Van Straaten et al.~(2004, 2005) have studied the
broad-band power spectra (including the noise components, the
low-frequency QPOs, and the kHz QPOs) of SAX J1808.4--3658 in detail
as well as the frequency correlations between the different
power-spectral components. Interestingly, using those frequency
correlations, van Straaten et al.~(2004, 2005) suggested that the
higher-frequency kHz QPO could also be identified during the 1998
outburst but at the lowest frequencies found so far in any kHz QPO
source (down to $\sim$150 Hz). Previous work (Wijnands \& van der Klis
1998b) on the aperiodic timing features of SAX J1808.4--3658 during
its 1998 outburst had already found these features but they could not
be identified as the higher-frequency kHz QPO due to their low
frequency and broad character.

Van Straaten et al.~(2004, 2005) also compared the results of SAX
J1808.4--3658 with those obtained for other non-pulsating neutron-star
LMXBs. In those other sources, the frequencies of the variability
components follow an universal scheme of correlations.  The
correlations observed for SAX J1808.4--3658 are similar but they show
a shift in the frequencies of the kHz QPOs. It is unclear what
physical mechanism(s) underlies this difference among sources (van
Straaten et al.~2004, 2005).

\begin{figure}[t]
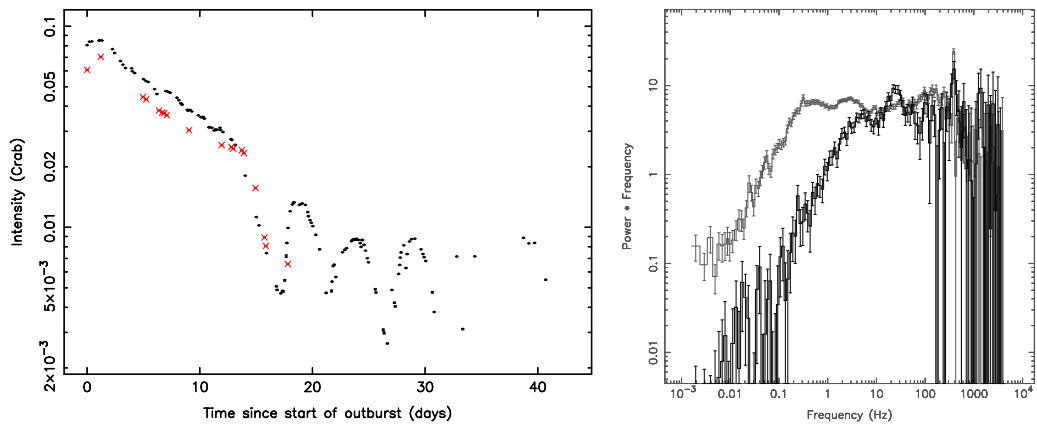

\begin{center}
\begin{tabular}{c}
\psfig{figure=rwijnands_fig8a.ps,height=5.5cm,angle=-90}~~~\psfig{figure=rwijnands_fig8b.ps,height=5.5cm}
\end{tabular}
\caption{
{\it Left}: The {\it RXTE}/PCA light curves of SAX J1808.4--3658
during its 1998 (crosses) and 2002 (dots) outbursts. The data were
taken from van Straaten et al.~(2005). {\it Right}: A comparison of
the power spectrum for SAX J1808.4--3658 obtained at roughly the same
flux levels ($\sim$0.044 Crab) during the 1998 outburst (light gray;
obtained on April 16, 1998) with that obtained during the 2002
outburst (black; October 23, 2002).
\label{fig:1808_1998v2002} }
\end{center}
\end{figure}

During the 1998 and 2002 outbursts of SAX J1808.4--3658, the source
exhibited similar X-ray fluxes. However, at similar flux levels, the
characteristic frequencies observed during the 1998 outburst are much
lower (by a factor of$\sim$10) than during the 2002 outburst (van
Straaten et al.~2005; see Fig.~\ref{fig:1808_1998v2002}). Again it is
unclear what causes this huge difference between the two outbursts but
it might be related to the 'parallel track' phenomena observed for the
kHz QPOs in the non-pulsating neutron-star LMXBs (e.g., van der Klis
2000).

\begin{figure}[t]
\begin{center}
\begin{tabular}{c}
\psfig{figure=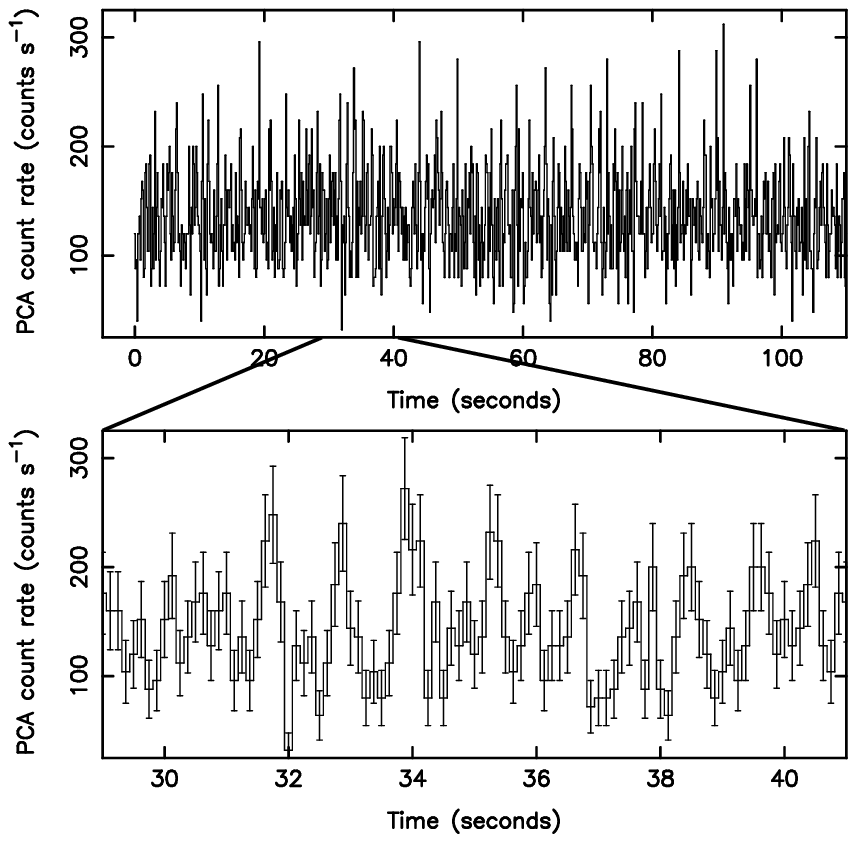,height=9.5cm,width=8cm}~~~~\psfig{figure=rwijnands_fig9b.ps,height=9.5cm}
\end{tabular}
\caption{ 
The violent 1 Hz flaring as observed during the 2000 and 2002
outbursts of SAX J1808.4-3658.  The flaring can be seen in the light
curve on the left panels and in the power spectra on the right panels.
\label{fig:1hz} }
\end{center}
\end{figure}

\subsubsection{The violent 1 Hz flaring
\label{subsubsection:1hzflaring}}

Violent flaring was observed on many occasions at a $\sim$1 Hz
repetition frequency during the late stages of the 2002 outburst
(Fig.~\ref{fig:1hz}), similar to what had been observed during the
2000 outburst. This proves that also this violent flaring is a
recurrent phenomenon and can likely be observed every time the source
is in this prolonged low-level activity state. Preliminary results
presented in Figure~\ref{fig:1hz} (right panels) show examples of
power spectra obtained during the end stages of the 2002
outburst. During certain observations the 1 Hz QPO is rather narrow
and its first overtone can be seen clearly (Fig.~\ref{fig:1hz} top
right panel). During other observations, the 1 Hz QPO is much broader
and its wings blend with the first overtone (Fig.~\ref{fig:1hz} middle
two right panels). In addition to the 1 Hz QPO, other QPOs around
30--40 Hz are sometimes seen (see also van Straaten et al.~2005). It
is unclear if this 30--40 Hz QPO is related to the low-frequency QPOs
discussed in \S~\ref{subsubsection:lfqpos} or if it is due to a
different mechanism. During certain observations the 1 Hz QPO becomes
very broad, turning into a band-limited noise component
(Fig.~\ref{fig:1hz} bottom right panel). The mechanism behind these
violent flares is not yet known and a detailed analysis of this
phenomenon is in progress.

\subsubsection{The pulsations
\label{subsubsection:pulsations}}

The pulsations could be detected at all flux levels with an amplitude
of 3\% -- 10\%.  There was no evidence for a 200.5 Hz subharmonic in
the data (upper limit of 0.38\% of the signal at 401 Hz; Wijnands et
al.~2003) confirming the interpretation of 401 Hz as the pulsar spin
frequency.  A detailed analysis and discussion of the coherent timing
analysis will be presented by Morgan et al.~(2005, in preparation).

\subsubsection{Observations at other wavelengths
\label{subsubsection:otherwavelengths}}

Rupen et al.~(2002b) reported the detection of SAX J1808.4--3658 at
radio wavelengths. On October 16, 2002, they found a 0.44-mJy source
at 8.5 GHz and a day later, the source was detected at 0.3 mJy. Monard
(2002) reported that on October 16, 2002, the optical counterpart was
detectable again at magnitudes similar to those observed at the peak
of the 1998 outburst.

\subsection{SAX J1808.4--3658 in quiescence
\label{subsection:1808q}}

In quiescence, SAX J1808.4--3658 has been observed on several
occasions with the {\it BeppoSAX} and {\it ASCA} satellites (Stella et
al.~2000; Dontani et al.~2000; Wijnands et al.~2002). The source was
very dim in quiescence, with a luminosity close to or lower than
$10^{32}$ ergs s$^{-1}$. Due to the low number of source photons
detected, these luminosities had large errors and no information could
be obtained on the spectral shape or possible variability in
quiescence. Due to the limited angular resolution of {\it BeppoSAX},
doubts were raised as to whether the source detected by this satellite
was truly SAX J1808.4--3658 or an unrelated field source (Wijnands et
al.~2002). Campana et al.~(2002) reported on a quiescent observation
of the source performed with {\it XMM-Newton} which resolved this
issue. They detected the source at a luminosity of $5 \times 10^{31}$
ergs s$^{-1}$ and found that the field around SAX J1808.4--3658 is
rather crowded with weak sources. Two such sources are relatively
close to SAX J1808.4--3658 and might have conceivably caused a
systematic positional offset during the {\it BeppoSAX}
observations. Despite this fact it is very likely that SAX
J1808.4--3658 was indeed detected during those {\it BeppoSAX}
observations.

Using {\it XMM-Newton}, Campana et al.~(2002) obtained enough photons
to extract a quiescent X-ray spectrum, which was not dominated by the
same thermal component seen in other quiescent neutron star
transients; such a thermal component is thought to be due to the
cooling of the neutron star in-between outbursts. However, the
spectrum of SAX J1808.4--3658 was dominated by a power-law shaped
component. The non-detection of the thermal component was used to
argue that the neutron star was anomalously cool, possibly due to
enhanced core cooling processes (Campana et al.~2002). It has been
argued (Stella et al.~2000; Campana et al.~2002) that the propeller
mechanism, which might explain (some of) the hard X-ray emission in
quiescence, is not likely to be active since this mechanism is
expected to stop operating at luminosities $<$$10^{33}$ ergs s$^{-1}$,
because at those luminosities the source should turn on as a radio
pulsar. Instead, it was proposed that the quiescent X-rays originate
in the shock between the wind of a turned-on radio pulsar and the
matter flowing out from the companion star (Stella et al.~2000;
Campana et al.~2002). Di Salvo \& Burderi (2003) suggested that the
quiescent X-rays could also be due to direct dipole radiation from the
radio pulsar. Using simple accretion disk physics and the quiescent
luminosity found by Campana et al.~(2002), they determined that the
magnetic field strength of the neutron star in SAX J1808.4--3658
should be in a quite narrow range of $(1 - 5) \times 10^8$ Gauss.

The quiescent optical counterpart of SAX J1808.4--3658 was studied by
Homer et al.~(2001). They reported that on August 10, 1999, the
orbital modulation was still present in white light observations
(estimated V magnitude of $\sim$20), with a semi-amplitude of
$\sim$6\%. It has the same phasing and approximately sinusoidal
modulation as seen during outburst, and with photometric minimum when
the pulsar is behind the companion star. During observations taken in
July 2000 the quiescent counterpart was even fainter and no
significant orbital modulation could be detected. Using these results,
it has been suggested that the optical properties of SAX J1808.4--3658
in quiescence are evidence of an active radio pulsar (Burderi et
al.~2003). Campana et al.~(2004) reported on the first optical
spectrum of this source during its quiescent state. They concluded
that a very high irradiating luminosity, a factor of $\sim$100 larger
than directly observed from the X-rays, must be present in the
systems, which was suggested to be derived from a rotation-powered
neutron star. If true, a pulsating radio source might be expected, but
a search at 1.4 GHz could not detected the source (Burgay et
al.~2003). This could be due to the effects of free-free absorption
and searches at higher frequencies to limit these effects might still
yield a pulsating radio source during the quiescent state of SAX
J1808.4-3658.

\section{XTE J1751--305
\label{section:1751}}

\subsection{The 2002 outburst
\label{subsection:1751_2002outburst}}

The second accretion-driven millisecond pulsar (XTE J1751--305) was
discovered on April 3, 2002 (Markwardt et al.~2002a). Its spin
frequency is 435 Hz (Fig.~\ref{fig:pulsations}) and the neutron star
is in a very small binary with an orbital period of only 42
minutes. The timing analysis of the pulsations gave a minimum mass for
the companion star of 0.013 solar mass and a pulse-frequency
derivative of $<$$3 \times 10^{-13}$ Hz s$^{-1}$. Assuming that the
mass transfer in this binary system was driven by gravitational
radiation, the distance toward the source could be constrained to at
least 7 kpc and the orbital inclination to 30$^\circ$--85$^\circ$,
resulting in a companion mass of 0.013--0.035 solar masses, suggesting
a heated helium dwarf (Markwardt et al.~2002a). {\it Chandra} briefly
observed the source, resulting in an arcsecond position (Markwardt et
al.~2002a).

The source reached a peak luminosity of $>$$2 \times 10^{37}$ ergs
s$^{-1}$, an order of magnitude brighter than the peak luminosity of
SAX J1808.4--3658. However, the outburst was very short with an
e-folding time of only $\sim$7 days (compared to $\sim$14 days for SAX
J1808.4--3658; Fig.~\ref{fig:lc}) resulting in a low outburst fluence
of only $\sim$$2.5 \times 10^{-3}$ ergs cm$^{-2}$ (Markwardt et
al.~2002a). A potential re-flare was seen two weeks after the end of
the outburst during which also a type-I X-ray burst was
seen. Preliminary analysis of the burst indicated that the burst did
not come from XTE J1751--305 but from another source in the field of
view. This was later confirmed (In 't Zand et al.~2003) and the burst
likely originated from the bright X-ray transient in Terzan 6. It was
also determined that the transient in Terzan 6 could not have produced
the re-flare (In 't Zand et al.~2003) suggesting that this re-flare
could still have come from XTE J1751--305. However, van Straaten et
al.~(2005) suggested (based on a X-ray color study using {\it
RXTE}/PCA observations) that this re-flare was emitted by one of the
background sources and not by XTE J1751--305. Van Straaten et
al.~(2005) also investigated the aperiodic timing properties of the
source (an example power spectrum is shown in Fig.~\ref{fig:lfqpos})
and the correlations between the characteristic frequencies of the
observed power-spectral components. The frequency correlations were
similar to those of the non-pulsating neutron-star LMXBs. In contrast
with the results obtained for SAX J1808.4--3658
(\S~\ref{subsubsection:lfqpos}), no frequency shift was required for
XTE J1751--305 to make the frequency correlations consistent with
those of the non-pulsating sources. Using these correlations, van
Straaten et al.~(2005) suggested that the highest-frequency noise
components in XTE J1751--305 are likely due to the same physical
mechanisms as the kHz QPOs. They also investigated the correlations
between the characteristic frequencies and the X-ray colors of the
source and concluded that it did not behave like an atoll source.

A previous outburst in June 1998 was discovered using archival {\it
RXTE}/ASM data (Markwardt et al.~2002a), suggesting a tentative
recurrence time of $\sim$3.8 years.  Miller et al.~(2003) reported on
high spectral resolution data of the source obtained with {\it
XMM-Newton} to search for line features in the X-ray
spectrum. However, they only detected a continuum spectrum dominated
by a hard power-law shaped component (power-law index of $\sim$1.44)
but with a 17\% contribution to the 0.5--10 keV flux from a soft
thermal (black-body) component with temperature of $\sim$1
keV. Gierlinski \& Poutanen (2004) studied in detail the X-ray
spectrum of the source during its 2002 outburst using the archival
{\it RXTE} and {\it XMM-Newton} observations. They find that XTE
J1751--305 exhibited very similar behavior as SAX J1808.4--3658 during
its 1998 outbursts. They also find that the pulse profile cannot be
described by a simple sinusoid, but that a second harmonic is needed
(the peak-to-peak amplitude of the fundamental was found to be 4.5\%
but of the second harmonic only 0.15\%). Gierlinski \& Poutanen (2004)
also report that a clear energy dependency of the pulse profile was
observed and that the higher energy photons arrive earlier than the
softer (this 'soft lag' reached $\sim$100 $\mu$s at about 10 keV,
where it saturated). Searches for the optical and near-infrared
counterparts were performed but no counterparts were found (Jonker et
al.~2003), likely due to the high reddening toward the source. These
non-detections did not constrain any models for the accretion disk or
possible donor stars.

\subsection{XTE J1751--305 in quiescence
\label{subsection:1751q}}

Recently, XTE J1751--305 was observed in quiescence using {\it
Chandra} (Wijnands et al.~2005). Sadly, they could not detect the
source in their $\sim$43 ksec observation, with 0.5--10 keV flux upper
limits between 0.2 and $2.7 \times 10^{-14}$ ergs s$^{-1}$ cm$^{-2}$
depending on assumed spectral shape, resulting in 0.5--10 keV
luminosity upper limits of $0.2 - 2 \times 10^{32}$ ($d$/ 8 kpc)$^2$
ergs s$^{-1}$, with $d$ the distance toward the source in kpc. Using
simple accretion disk physics in combination with these luminosity
upper limits, Wijnands et al.~(2005) could constrain the magnetic
field of the neutron star in this system to be less than $3 - 7 \times
10^8 {d \over {\rm 8~kpc}}$ Gauss (depending on assumed spectral shape
of the quiescent spectrum).

\section{XTE J0929--314
\label{section:0929}}

\subsection{The 2002 outburst
\label{subsection:0929_2002outburst}}

The third accretion-driven millisecond X-ray pulsar XTE J0929--314 had
already been detected with the {\it RXTE}/ASM on April 13, 2002
(Remillard 2002) but was only found to be harboring a millisecond
pulsar with a pulsations frequency of 185 Hz
(Fig.~\ref{fig:pulsations}) on May 2nd when observations of the source
were made using the {\it RXTE}/PCA (Remillard et al.~2002). Galloway
et al.~(2002) reported on the detection of the 44-min orbital period
of the system which is remarkably similar to that of XTE J1751--305. A
minimum mass of 0.008 solar mass was obtained for the companion star
and a pulse-frequency derivative of $(-9.2 \pm 0.4) \times 10^{-14}$
Hz s$^{-1}$. Galloway et al.~(2002) suggested that this spin down
torque may arise from magnetic coupling to the accretion disk, a
magneto-hydrodynamic wind, or gravitational radiation from the rapidly
spinning neutron star. Assuming gravitational radiation as the driving
force behind the mass transfer, Galloway et al.~(2002) found a lower
limit to the distance of 6 kpc. They also reported on the detection of
a QPO at 1 Hz (Fig.~\ref{fig:lfqpos}). Full details of this QPO and
the other aperiodic power-spectral components are presented by van
Straaten et al.~(2005). Just as they found for SAX J1808.4--3658, the
frequency correlations for XTE J0929--314 were similar to those
observed for the non-pulsating sources but with an offset in the
frequencies of the highest-frequency components. These correlations
allowed van Straaten et al.~(2005) to identify those components as
related to the kHz QPOs. Studying the correlated spectral and timing
variability, they concluded that the behavior of XTE J0929--314 was
consistent with that of an atoll source.

Juett et al.~(2003) obtained high resolution spectral data using the
{\it Chandra} gratings. Again the spectrum is well fitted by a
power-law plus a black body component, with a power-law index of 1.55
and a temperature of 0.65 keV. Similar to XTE J1751--305, no emission
or absorption features were found. No orbital modulation of the X-ray
flux was found implying an upper limit on the inclination of
85$^\circ$. Greenhill et al.~(2002) reported the discovery of the
optical counterpart of the system with a V magnitude of 18.8 on May
1st, 2002 (see also Cacella 2002). Castro-Tirado et al.~(2002)
obtained optical spectra of the source on May 6--8 in the range
350--800 nm and found emission lines from the C III - N III blend and
H-alpha, which were superposed on a blue continuum. These optical
properties are typical of X-ray transients during outburst. Rupen et
al.~(2002a) discovered the radio counterpart of the source using the
VLA with 4.86 GHz flux of 0.3--0.4 mJy.

\subsection{XTE J0929--314 in quiescence
\label{subsection:0929q}}

Recently, Wijnands et al.~(2005) also observed XTE J0929--314 in its
quiescent state with {\it Chandra}. For this source, they detected 22
source photons (in the energy range 0.3--8 keV) in $\sim$24.4 ksec of
on-source time. This small number of photons detected did not allow
for a detailed spectral analysis of the quiescent spectrum, but they
could demonstrate that the spectrum is harder than a simple thermal
emission (which might have been due to the cooling neutron star that
has been heated during outbursts). Assuming a power-law spectral model
for the time-averaged (averaged over the whole observation) X-ray
spectrum, they obtained a power-law index of $\sim$1.8 and an
unabsorbed X-ray flux of $\sim$$6 \times 10^{-15}$ ergs s$^{-1}$
cm$^{-2}$ (for the energy range 0.5--10 keV), resulting in a 0.5--10
keV X-ray luminosity of $\sim$$7\times 10^{31}$ ($d$/10 kpc)$^2$ ergs
s$^{-1}$, with $d$ the distance in kpc. The thermal component usually
seen in quiescent neutron star LMXBs could not be detected, with a
maximum contribution to the 0.5--10 keV flux of $\sim$30\%. Wijnands
et al.~(2005) also found that the quiescent count rate of XTE
J0929--314 was variable at the 95\% confidence level, but no
conclusive evidence was found for associated spectral variability. The
properties of XTE J0929--314 in its quiescent state are remarkably
similar to that observed for SAX J1808.4--3658
(\S~\ref{subsection:1808q}) which might suggest that such behavior is
common among accretion-driven millisecond X-ray pulsars. However,
recent work on several other weak quiescent neutron-star X-ray
binaries (e.g., Jonker et al.~2004a, b ; Tomsick et al.~2004), which
do not exhibit pulsations during their X-ray outbursts, suggests that
also such systems can resemble SAX J1808.4--3658 during quiescent
(i.e., they could be almost as faint and hard as SAX J1808.4--3658 in
quiescence; see Wijnands et al.~2005 for a in-dept discussion).
Similar to what they did for XTE J1751--305, Wijnands et al.~(2005)
could constrain the neutron-star magnetic field strength in XTE
J0929--314 to be $<$$3\times 10^9 {d\over {\rm 10~kpc}}$ Gauss.

\section{XTE J1807--294
\label{section:1807}}

\begin{figure}[t]
\begin{center}
\begin{tabular}{c}
\psfig{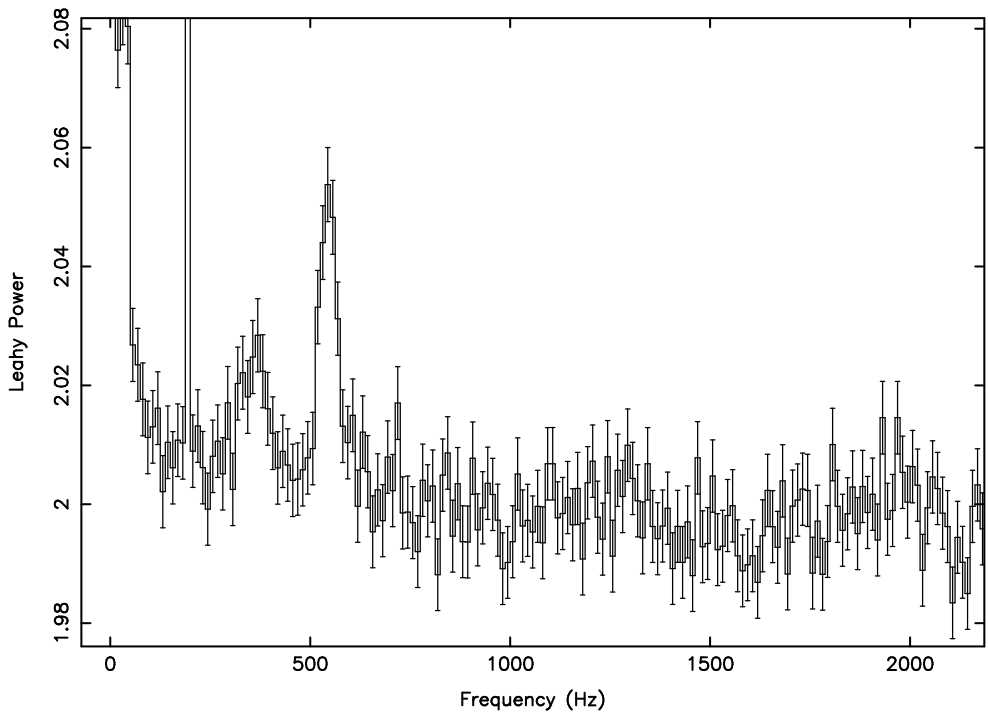}
\end{tabular}
\caption{ 
The power spectrum of XTE J1807--294 as obtained using {\it
RXTE}/PCA. The two simultaneous kHz QPOs are clearly visible.
\label{fig:khz1807} }
\end{center}
\end{figure}

The fourth millisecond X-ray pulsar XTE J1807--294 with a frequency of
191 Hz, was discovered on February 21, 2003 (Markwardt et al.~2003a;
Fig.~\ref{fig:pulsations}). The peak flux was only 58 mCrab (2--10
keV, measured on February 21, 2003). The orbital period was determined
(Markwardt et al.~2003c) to be $\sim$40 minutes making it the shortest
orbital period of all accretion-driven millisecond pulsars now
known. Markwardt et al.~(2003c) reported the best known position of
the source based on a {\it Chandra} observation. Using the {\it
RXTE}/PCA data, kHz QPOs have been detected for this system and it was
found that the frequency separation between the two kHz QPOs was
consistent with being equal to the neutron-star spin frequency
(Markwardt et al.~2005 in preparation; Fig.~\ref{fig:khz1807}). This
makes XTE J1807--294 consistent with the classification of Wijnands et
al.~(2003) of the neutron-star LMXBs into 'fast' and 'slow' rotators,
with XTE J1807--294 a slow rotator. A detailed analysis of the
correlations between the kHz QPOs and the low-frequency features (see
Fig.~\ref{fig:lfqpos}) in this source will be reported by Linares et
al.~(2005, in preparation). The preliminary results of that analysis
shows that also for XTE J1807--294 the frequency correlations are
similar to those observed for the non-pulsating sources but with an
offset in the frequencies of the highest-frequency components (similar
to what was found for SAX J1808.4--3658 and possible XTE J0929--314;
Van Straaten et al.~2005).  Campana et al.~(2003) reported on a {\it
XMM-Newton} observation of this source taken on March 22,
2003. Assuming a distance of 8 kpc, the 0.5--10 keV luminosity during
that observation was $2 \times 10^{36}$ ergs s$^{-1}$. They could
detect the pulsations during this observation with a pulsed fraction
of 5.8\% in the 0.3--10 keV band (increasing with energy) and a nearly
sinusoidal pulse profile. Furthermore, using the same data Kirsch et
al.~(2004; see also Kirsch \& Kendziorra 2003) reported on the mass
function of this system and found a minimal mass for the companion
star of 0.007 $M_\odot$ when assuming a canonical neutron star mass of
1.4 $M_\odot$. The spectral data are well fit by a continuum model,
assumed to be an absorbed Comptonisation model plus a soft
component. The latter component only contributed 13\% to the
flux. Again no emission or absorption lines were found. No detections
of the counterparts of the system at other wavelengths have been
reported so far.

\section{XTE J1814--338
\label{section:1814}}

The fifth system (XTE J1814--338) was discovered on June 5, 2003 and
has a pulse frequency of 314 Hz (Markwardt et al.~2003b;
Fig.~\ref{fig:pulsations}), with an orbital period of 4.3 hr and a
minimum companion mass of 0.15 solar mass (Markwardt et
al.~2003d). This 4.3 hr orbital period makes it the widest binary
system among the accretion-driven millisecond pulsars and also the one
most similar to the general population of low-luminosity neutron star
LMXBs (the atoll sources). XTE J1814--338 exhibited many type-I X-ray
bursts, which showed burst oscillations with a frequency consistent
with the neutron star spin frequency (Markwardt et al.~2003d,
Strohmayer et al.~2003). A distance of $\sim$8 kpc was obtained from
the only burst which likely reached the Eddington luminosity. The
burst oscillations are strongly frequency- and phase-locked to the
persistent pulsations (as was also seen for SAX J1808.4--3658;
Chakrabarty et al.~2003) and two bursts showed evidence of a frequency
decrease of a few tenths of a Hz during the onset of the burst,
suggesting a spin down. Strohmayer et al.~(2003) also reported on the
detection of the first harmonic of the burst oscillations, which is
the first time that this has been seen in any burst-oscillation
source. This harmonic could arise from two hot-spots on the surface,
but Strohmayer et al.~(2003) suggested that if the burst oscillations
arise from a single bright region, the strength of the harmonic would
suggest that the burst emission is beamed (possibly due to a stronger
magnetic field strength than in non-pulsating LMXBs). Bhattacharyya et
al.~(2004) used the non-sinusoidal burst oscillation light curves to
constrain the parameters of the neutron star in XTE J1814--338; they
obtained a dimensionless radius to mass ratio of $Rc^2/GM = 3.9 - 4.9$
for the neutron star in this source. They find that the bursting hot
spot on the neutron-star surface remains always large, with an angular
radius $>$$25^\circ$. Their study also suggest that the inclination of
the source is greater than 50$^\circ$ and that the secondary companion
is a hydrogen main sequence star that is significantly bloated
(possibly due to X-ray heating).

Wijnands \& Homan (2003) analyzed the {\it RXTE}/PCA data of the
source obtained between June 8 and 11, 2003. The overall shape of the
3--60 keV power spectrum is dominated by a strong broad band-limited
noise component (Fig.~\ref{fig:lfqpos}), which could be fitted by a
broken power-law model with a broad bump superimposed on it at
frequencies above the break frequency. These characteristics make the
power spectrum of XTE J1814--338 very similar to that observed in the
non-pulsing low-luminosity neutron-star LMXBs (the atoll sources) when
they are observed at relatively low X-ray luminosities (i.e., in the
so-called island state). This is consistent with the hard power-law
X-ray spectrum of the source reported by Markwardt et
al.~(2003d). This resemblance of XTE J1814--338 to the atoll sources
was further strengthened (Wijnands \& Homan 2003) by the fact that the
source is consistent with the relation between the break frequency and
the frequency of the bump found for atoll sources by Wijnands \& van
der Klis (1999). Van Straaten et al.~(2005) performed an in-depth
analysis of all publicly available {\it RXTE}/PCA data of XTE
J1814--338 to study the power-spectral components and the correlations
between their characteristic frequencies. Using those correlations and
by comparing them to other sources, they could identify several
components that are related to the kHz QPOs. They also found that the
frequency correlations were identical to the non-pulsating sources
with no need for a frequency shift. This is similar to what they found
for XTE J1751--305 but different from SAX J1808.4--3658 and XTE
J0929--314 (and for XTE J1807--294 as found by Linares et al.~2005, in
preparation). The reason(s) for this difference between accreting
millisecond X-ray pulsars is not know (see Van Straaten et al.~2005
for an extended discussion). From the correlations between the
spectral and timing variability it was confirmed that the behavior of
XTE J1814--338 was consistent with that of an atoll sources (van
Straaten et al.~2005; see also Wijnands \& Homan 2003).

Wijnands \& Reynolds (2003) reported that the position of XTE
J1814--338 was consistent with the {\it EXOSAT} slew source EXMS
B1810--337 which was detected on September 2nd, 1984. If this
identification is correct, then its recurrence time can be inferred to
be less than 19 years but more than 5 years (the time since the {\it
RXTE}/PCA bulge scan observations started in February 1999), unless
the recurrence time of the source varies significantly. Krauss et
al.~(2003) reported the best position of the source based on a {\it
Chandra} observation and on the detection of the likely optical
counterpart of the source (with magnitudes of B = 17.3 and R = 18.8 on
June 6). Steeghs (2003) reported on optical spectroscopy of this
possible counterpart, specifically on the discovery of prominent
hydrogen and helium emission lines, confirming the connection between
the optical source and XTE J1814--338.

\section{IGR J00291+5934}

Very recently, on December 2, 2004, the European Gamma-ray satellite
{\it INTEGRAL} discovered a new X-ray transient named IGR J00291+5934
(Eckert et al.~2004). A day later, {\it RXTE} observed the source and
it was found that this source harbors a 598.88 Hz accretion-driven
millisecond X-ray pulsar (Markwardt et al.~2004a; see Galloway et
al.~2005 for the analyze of all data obtained for this source). The
pulsed amplitude was approximately 6\% with no evidence for
harmonics. The amplitude decreased with increasing photon energy and
the soft photons arrive later than the hard ones (by up to 85 $\mu$s;
Galloway et al.~2005) The X-ray spectrum could be fitted with an
absorbed power-law model with photon index of 1.7 and a column density
of $7\times 10^{21}$ cm$^{-2}$.  In Figure~\ref{fig:lfqpos} the power
spectrum between 0.001 and 10,000 Hz is show of the source, clearly
showing significant aperiodic variability (see also Markwardt et
al.~2004a). Interestingly, of all six accreting millisecond X-ray
pulsars the break in the power spectrum is at the lowest frequency for
IGR J00291+5934: during the peak of the outburst the break frequency
for this source was $\sim$0.01 Hz compared to $>$0.1 Hz for the other
sources at their outburst peaks.  Markwardt et al.~(2004b) used
additional {\it RXTE} observations to determine that the orbital
period of the system was $\sim$2.45 hours and they obtained a mass
function of $(2.81 \pm 0.02) \times 10^{-5}$ M$_\odot$ (see also
Galloway et al.~2005). For a neutron star mass of 1.4 M$_\odot$ this
results in a lower limit on the mass for the companion star of 0.038
M$_\odot$, possible a brown dwarf (Galloway et al.~2005). The source
reached its peak flux (29 mCrab; 2--10 keV) on December 3, 2004, after
which the fluxes decreased in a linear way, until around December 11,
2004, when the rate of decline had increased slightly (Swank \&
Markwardt 2004). On December 14, 2004, {\it Chandra} performed a brief
($\sim$18 ksec) observation of the source using the ACIS-S/HETG
combination. Nowak et al.~(2004) reported that the source was at a
flux level of $\sim$1 mCrab and its X-ray spectrum could be
well-fitted by an absorbed power-law model with a column density of
$\sim$$3\times 10^{21}$ cm$^{-2}$ and a photon index of $\sim$1.9. A
possible iron line feature was also reported by the same authors. The
flux value confirmed the steady decline as seen by Swank
\& Markwardt (2004) and soon after this the source could not be
detected anymore with {\it RXTE}/PCA (see Galloway et al.~2005 for the
full X-ray light curve of the source during this outburst). No X-ray
bursts, dips, eclipses, or kHz QPOs (with upper limits of $\sim$1\%
rms) were found (Galloway et al.~2005).

\begin{figure}[t]
\begin{center}
\begin{tabular}{c}
\psfig{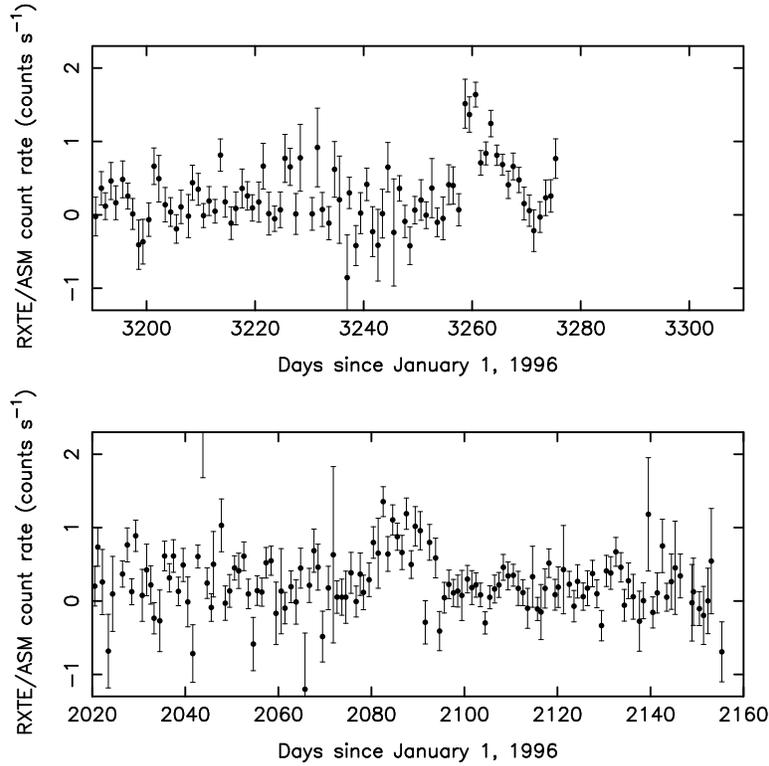}
\end{tabular}
\caption{ 
The {\it RXTE}/ASM light curve of IGR J00291+5934 during the most
recent (December 2004) outburst (top panel; near day 3260) and its
likely September 2001 outburst (bottom panel; near day 2080).  These
light curves were made using the public ASM data available at
http://xte.mit.edu/ASM\_lc.html. The count rates are for the 2--12 keV
energy range and are daily averages.
\label{fig:0029_ASM} }
\end{center}
\end{figure}

An optical counterpart was proposed by Fox \& Kulkarni (2004; see this
reference for a finding chart) with an R magnitude during outburst of
$\sim$17.4. The detection of broad emission lines of HeII and
H$_\alpha$ from this tentative optical counterpart strongly support
this identification (Roelofs et al.~2004; see also Filippenko et
al.~2004). Pooley (2004) found a 1.1 mJy radio source (15 GHz) at a
position consistent with that of the optical counterpart which likely
is the radio counterpart of the source which faded during the outburst
(Fender et al.~2004, who also detected the source at 5 GHz at a flux
of $\sim$250 $\mu$Jy), although the decay did not seem to be very
rapid (Rupen et al.~2004). Steeghs et al.~(2004) reported the
detection of a decaying infrared counterpart of the source.

After the discovery of this new source, Remillard (2004) constructed a
mission-long light curve using the {\it RXTE}/ASM data and found that
in November 26--28, 1998, and in September 11--21, 2001, the source
might have exhibited other outbursts (see Fig.~\ref{fig:0029_ASM} for
the {\it RXTE}/ASM light curve of the December 2004 and September 2001
outbursts of the source). If confirmed this would give a recurrence
time for the outburst of approximately 3 years. In 't Zand \& Heise
(2004) did not detected the source with the WFCs aboard {\it BeppoSAX}
during a net exposure time of 2.9 Msec on the source. The {\it
RXTE}/ASM detections reported by Remillard (2004) were not covered;
the first WFCs data were obtained 16 days and 11 days after the two
possible outbursts, respectively, and it is well possible that the
source had by then decayed below the sensitivity limit of the WFCs.

\section{Theoretical work
\label{section:theory}}

This chapter is intended to be an observational overview, thus I will
not go into detail on the theoretical papers published on
accretion-driven millisecond pulsars. Instead, I will briefly list
some of those papers, most of which focus on SAX J1808.4--3658 since
the other five systems have only been found recently (with IGR
J00291+5934 discovered very recently). Since the discovery of SAX
J1808.4--3658, several studies have tried to constrain the properties
(i.e., radius, mass, magnetic field strength) of the neutron star in
this system (Burderi \& King 1998; Psaltis \& Chakrabarty 1999;
Bhattacharyya 2001), while others proposed that the compact object is
not a neutron star at all, but instead a strange star (see, e.g., Li
et al.~1999; Datta et al.~2000; Zdunik et al.~2000 and references
there-in). Other studies focused on the evolutionary history of SAX
J1808.4--3658 (Ergma \& Antipova 1999) or on the nature of the
companion star (Bildsten \& Chakrabarty 2001, who suggested a brown
dwarf companion star).  Recently, Nelson \& Rappaport (2003)
investigated the evolutionary history of ultracompact binaries such as
XTE J0929--314 and XTE J1751--305 (they focused on those two systems
but likely their conclusions can also be applied to XTE
J1807--294). Rappaport et al.~(2004) investigated how accretion could
occur in millisecond X-ray pulsars and they postulated that those
systems can continue to accrete from a thin disk, even for accretion
rates that place the magnetospheric radius well beyond the co-rotation
radius.

The discovery of the accretion-driven millisecond pulsars raises an
important question: why are those systems different from other neutron
star LMXBs for which no pulsations have been found. Cumming et
al.~(2001; see also Rai Choudhuri \& Konar 2002; Cumming 2004)
suggested that the low time-averaged accretion rate of SAX
J1808.4--3658 might explain why this source is a pulsar. Although the
remaining four pulsars were not know at the time of writing of that
paper, the same arguments can be used for those systems: when the
time-averaged accretion rate is sufficiently high, the neutron star
magnetic field might be buried by the accreted matter and does not
have time to dissipate through the accreted material. However, for the
millisecond X-ray pulsars the time-averaged accretion rate is
sufficiently low that the magnetic field dissipation can indeed
happen, giving those systems a magnetic field still strong enough to
disturb the flow of the accreted matter. However, more neutron-star
LMXBs with low time-averaged accretion rate must be found and studied
in detail to verify that they all indeed harbor a millisecond
pulsar. If an exception is found, the screening model might only be
part of the explanation and alternative ideas need to be explored
(see, e.g., Titarchuk et al.~2002).

\section{Conclusion
\label{section:conclusions}}

From this review it is clear that {\it RXTE} has played a vital role
in the discovery and study of accretion-driven millisecond X-ray
pulsars. The detailed studies performed with {\it RXTE} for those
systems have yielded break-throughs in our understanding of kHz QPOs
and burst oscillations. Furthermore, three of these accreting pulsars
are in ultrashort binaries which will constrain evolutionary paths for
this type of systems (e.g., see Nelson \& Rappaport 2003). However, it
is also clear that the six systems do not form a homogeneous group;
their pulsation frequencies span the range between 185 Hz and 599 Hz
(Fig.~\ref{fig:pulsations}), their orbital periods fall between 40
minutes and 4.3 hours, their X-ray light curves are very different
(Fig.~\ref{fig:lc}), and also their aperiodic variability properties
(Fig.~\ref{fig:lfqpos}). More, well studied outbursts of the currently
known systems are needed as well as discoveries of additional
systems. At the moment only {\it RXTE} is capable of performing the
necessary timing observations. After {\it RXTE}, an instrument with
similar or better capabilities is highly desirable for our
understanding of accretion-driven millisecond pulsars and their
connection with the non-pulsating neutron-star LMXBs.

\section*{Acknowledgments}

I thank Stefanie Wachter for kindly providing the optical data used in
Figure~\ref{fig:2000_lc}.

\section{References}

\parindent=0pt

Bhattacharya, D., \& van den Heuvel, E.P.J.,  1991, {\it Physics
Reports}, 203, 1--124
 
Bhattacharyya, S., 2001, {\it ApJ}, 554, L185--L188
 
Bhattacharyya, S. et al., 2004, {\it ApJ}, in press
 (astro-ph/0402534)
 
Bildsten, L. \& Chakrabarty, D., 2001, {\it ApJ}, 557, 292--296
 
Burderi, L. \& King. A.R., 1998, {\it ApJ}, 505, L135--L137
 
Burderi, L. et al., 2003, {\it A\&A}, 404, L43--L46
 
Burgay, M., et al., 2003, {\it ApJ}, 589, 902--910
 
Cacella, P., 2002, {\it IAUC}, 7893
 
Campana, S. et al., 2002, {\it ApJ}, 575, L15--L19
 
Campana, S. et al., 2003, {\it ApJ}, 594, L39--L42
 
Campana, S. et al., 2004, {\it ApJ}, 614, L49--L52
 
Castro-Tirado, A.J. et al., 2002, {\it IAUC}, 7895
 
Chakrabarty, D. 2004, To appear in {\it Binary Radio Pulsars},
F. A. Rasio \& I. H. Stairs (eds.), ASP Conf. Ser. (astro-ph/0408004)
 
Chakrabarty, D. \& Morgan, E. H., 1998, {\it Nature}, 394, 346--348
 
Chakrabarty, D. et al., 2003, {\it Nature}, 424, 42--44
 
Cui, W. et al., 1998, {\it ApJ}, 504, L27--L30
 
Cumming, A. 2004, To appear in {\it Binary Radio Pulsars} F. A. Rasio \&
I. H. Stairs (eds.), ASP Conf. Ser. (astro-ph/0404518)
 
Cumming, A. et al., 2001, {\it ApJ}, 557, 958--966
 
Datta, B. et al., 2000, {\it A\&A}, 355, L19--L22
 
Di Salvo, T. \& Burderi, L., 2003, {\it A\&A}, 397, 723--727
 
Dotani, T. et al., 2000, {\it ApJ}, 543, L145--L148
 
Eckert, D. et al., 2004, {\it ATEL}, 352
 
Ergma, E. \& Antipova, J., 1999, {\it A\&A}, 343, L45--L48
 
Fender, R. et al., 2004, {\it ATEL}, 361
 
Filippenko, A. V. et al., 2004, {\it ATEL}, 366
 
Ford, E.C., 2000, {\it ApJ}, 535, L119--L122
 
Fox, D. B. \& Kulkarni S. R., 2004, {\it ATEL}, 354
 
Gaensler, B.M. et al., 1999, {\it ApJ}, 522, L117--L119
 
Galloway, D.K. et al., 2002, {\it ApJ}, 576, L137--L140
 
Galloway, D.K. et al., 2005, ApJL, submitted (astro-ph/0501064)

Gierlinski, M. \& Poutanen, J., 2004, {\it MNRAS}, submitted
(astro-ph/0411716)
 
Gierlinski, M. et al., 2002, {\it MNRAS}, 331, 141--153
 
Giles, A.B. et al., 1998, {\it IAUC}, 6886
 
Giles, A.B. et al., 1999, {\it MNRAS}, 304, 47--51
 
Gilfanov, M. et al., 1998, {\it A\&A}, 338, L83--L86
 
Greenhill, J.G. et al., 2002, {\it IAUC}, 7889
 
Heindl, W.A. \& Smith, D.M., 1998, {\it ApJ}, 506, L35--L38
 
Heise, J. et al., 1999, {\it ApL\&C}, 38, 297--300
 
Homer, L. et al., 2001, {\it MNRAS}, 325, 1471--1476
 
Jonker, P.G. et al., 2000, {\it ApJ}, 540, L29--L32
 
Jonker, P.G. et al., 2003, {\it MNRAS}, 344, 201--206
 
Jonker, P.G. et al., 2004a, {\it MNRAS}, 349, 94--98

Jonker, P.G. et al., 2004b, {\it MNRAS}, 354, 666--674

Juett, A.M. et al., 2003, {\it ApJ}, 587, 754--760
 
In 't Zand, J. \& Heise, J., 2004, {\it ATEL}, 362
 
In 't Zand, J.J.M. et al., 1998, {\it A\&A}, 331, L25--L28
 
In 't Zand, J.J.M. et al., 2001, {\it A\&A}, 372, 916--921
 
In 't Zand, J.J.M. et al., 2003, {\it A\&A}, 409, 659--663
 
Kato, S., 2004, {\it PASJ}, 56, 905--922
 
Kazarovets, E.V. et al., 2000, {\it IBVS}, 4870
 
Kirsch, M.G.F. \& Kendziorra, E., 2003, {\it ATEL}, 148
 
Kirsch, M. G. F. et al., 2004, {\it A\&A}, 423, L9--L12
 
Kluzniak, W. et al., 2004, {\it ApJ}, 603, L89--L92
 
Krauss, M.I. et al., 2003, {\it IAUC}, 8154
 
Lamb, F.K. \& Miller, M.C., 2004, {\it ApJ Letters}, submitted
(astro-ph/0308179)
 
Lee, W. H., et al., 2004, {\it ApJ}, 603, L93--L96
 
Li, X.-D. et al., 1999, {\it Physical Review Letters}, 83, 3776--3779
 
Markwardt, C.B. et al., 2002a, {\it ApJ}, 575, L21--L24
 
Markwardt, C.B. et al., 2002b, {\it IAUC}, 7993
 
Markwardt, C.B. et al., 2003a, {\it IAUC}, 8080
 
Markwardt, C.B. et al., 2003b, {\it IAUC}, 8144
 
Markwardt, C.B. et al., 2003c, {\it ATEL}, 127
 
Markwardt, C.B. et al., 2003d, {\it ATEL}, 164
 
Markwardt, C.B. et al., 2004a, {\it ATEL}, 353
 
Markwardt, C.B. et al., 2004b, {\it ATEL}, 360
 
Marshall, F.E., 1998, {\it IAUC}, 6876
 
Miller, J.M. et al., 2003, {\it ApJ}, 583, L99--L102
 
Monard, B., 2002, {\it VSNet alert}, 7550
 
Nelson, L. A. \& Rappaport, S., 2003, {\it ApJ}, 598, 431--445
 
Nowak, M. A. et al., 2004, {\it ATEL}, 369
 
Pooley, G., 2004, {\it ATEL}, 355
 
Poutanen, J. \& Gierlinski, M., 2003, {\it MNRAS}, 343, 1301--1311
 
Psaltis, D. \& Chakrabarty, D., 1999, {\it ApJ}, 521, 332--340
 
Rai Choudhuri, A. \& Konar, S., 2002, {\it MNRAS}, 332, 933--944
 
Rappaport, S. A. et al.,  2004, {\it ApJ}, 606, 436--443
 
Remillard, R.A., 2002, {\it IAUC}, 7888
 
Remillard, R.A. et al., 2002, {\it IAUC}, 7893
 
Remillard, R., 2004, {\it ATEL}, 357
 
Revnivtsev, M.G., 2003, {\it AstL}, 29, 383--386
 
Roche, P. et al., 1998, {\it IAUC}, 6885
 
Roelofs, G. et al., 2004, {\it  ATEL}, 356
 
Rupen, M.P. et al., 2002a, {\it IAUC}, 7893
 
Rupen, M.P. et al., 2002b, {\it IAUC}, 7997
 
Rupen, M. et al.,  2004, {\it ATEL}, 364
 
Steeghs, D., 2003, {\it IAUC}, 8155
 
Steeghs, D. et al.,  2004, {\it ATEL}, 363
 
Stella, L. et al., 2000, {\it ApJ}, 537, L115--L118
 
Strohmayer, T. \& Bildsten, L., 2003, To appear in {\it Compact
Stellar X-ray sources}, W.H.G. Lewin \& M. van der Klis (eds.),
Cambridge University Press (astro-ph/0301544)
 
Strohmayer, T.E. et al., 1996, {\it ApJ}, 469, L9--L12
 
Strohmayer, T.E. et al., 2003, {\it ApJ}, 596, L67--L70
 
Swank, J.H. \& Markwardt, C.B., 2004, {\it ATEL}, 365
 
Titarchuk, L. et al., 2002, {\it ApJ}, 576, L49--L52

Tomsick, J.A. et al., 2004, {\it ApJ}, 610, 933--940
 
Vaughan, B.A. et al., 1994, {\it ApJ}, 435, 362--371

Van der Klis, M., 2000, {\it ARA\&A}, 38, 717--760
 
Van der Klis, M., 2004, To appear in {\it Compact Stellar X-ray
sources}, W.H.G. Lewin \& van der Klis (eds.), Cambridge University
Press (astro-ph/0410551)

Van der Klis, M. et al., 1996, {\it ApJ}, 469, L1--L4
 
Van der Klis, M. et al., 2000, {\it IAUC}, 7358

Van Straaten, S. et al., 2004, In {\it The Restless High-Energy
Universe}, eds. E.P.J. van den Heuvel, J.J.M. in 't Zand, \&
R.A.M.J. Wijers (Elsevier), {\it Nuclear Physics B}, 132, 664--667
 
Van Straaten, S. et al., 2005, {\it ApJ}, in press (astro-ph/0410505)
 
Wachter, S. \& Hoard, D.W., 2000, {\it IAUC}, 7363
 
Wachter, S. et al., 2000, {\it HEAD}, 32, 24.15
 
Wang, Z. et al., 2001, {\it ApJ}, 563, L61--L64
 
Wijnands, R., 2003, {\it ApJ}, 588, 425--429
 
Wijnands, R., 2004a, In {\it The Restless High-Energy Universe},
eds. E.P.J. van den Heuvel, J.J.M. in 't Zand, \& R.A.M.J. Wijers
(Elsevier), {\it Nuclear Physics B}, 132, 496--505

Wijnands, R., 2004b, In {\it X-ray Timing 2003: Rossi and Beyond},
eds. P. Kaaret, F.K. Lamb, \& J.H. Swank (Melville, NY), {\it AIP},
714, 209--216
 
Wijnands, R. \& van der Klis, M., 1998a, {\it Nature}, 394, 344--346
 
Wijnands, R. \& van der Klis, M., 1998b, {\it ApJ}, 507, L63--L66
 
Wijnands, R. \& van der Klis, M., 1999, {\it ApJ}, 514, 939--944
 
Wijnands, R. \& Homan, J., 2003, {\it ATEL}, 165
 
Wijnands, R. \& Reynolds, A., 2003, {\it ATEL}, 166
 
Wijnands, R. et al., 2001, {\it ApJ}, 560, 892--896
 
Wijnands, R. et al., 2002, {\it ApJ}, 571, 429--434
 
Wijnands, R. et al., 2003, {\it Nature}, 424, 44--47
 
Wijnands, R. et al., 2005, {\it ApJ}, in press (astro-ph/0406057)
 
Zdunik, J.L. et al., 2000, {\it A\&A}, 359, 143--147

\end{document}